\documentclass[aps, prd, showpacs,amsfonts,superscriptaddress, nofootinbib, showkeys, notitlepage, 11pt]{revtex4-1}

\usepackage[hyperindex]{hyperref}

\usepackage{graphicx}
\usepackage{array}
\usepackage{amsmath}

\newcommand{\mtiny}[1]{{\mbox{\tiny #1}}}

\newcommand{\x}{x}
\newcommand{\xp}{x'}
\newcommand{\dd}{\delta^{(3)}(\x, \xp)}
\newcommand{\ddn}{\delta^{(3)}}
\newcommand{\ddc}{\ddn}

\newcommand{\bs}[1]{\mathbf{#1}}
\newcommand{\3}{\, {}^{\mbox{\tiny (3)}}\!}

\newcommand{\ff}{{\bs{\bar{f}}}}
\newcommand{\xixi}{{\boldsymbol{\bar{\xi}} }}
\newcommand{\aaa}{{\boldsymbol{\bar{a}}}}
\newcommand{\nunu}{{\boldsymbol{\bar{\nu}}}}

\newcommand{\bsnu}{\boldsymbol{\nu}}
\newcommand{\bsomega}{\boldsymbol{\omega}}

\begin{document}

\author{Davi C. Rodrigues}\email{davi.rodrigues@cosmo-ufes.org}\affiliation{\footnotesize Center for Astrophysics and Cosmology \& Department of Physics, CCE, Federal University of Esp\'irito Santo, Av.~Fernando Ferrari 514, Vit\'{o}ria, ES, Brazil}\author{Mariniel Galv\~ao}\email{mariniel.galvao@cosmo-ufes.org}\affiliation{\footnotesize Center for Astrophysics and Cosmology \& Department of Physics, CCE, Federal University of Esp\'irito Santo, Av.~Fernando Ferrari 514, Vit\'{o}ria, ES, Brazil}\author{Nelson Pinto-Neto}\email{nelsonpn@cbpf.br}\affiliation{\footnotesize COSMO -– Centro Brasileiro de Pesquisas F\'{\i}sicas, R. Xavier Sigaud, 150, Urca, Rio de Janeiro, Brazil}

\title{Hamiltonian analysis of General Relativity and extended gravity from the iterative Faddeev-Jackiw symplectic approach}

\begin{abstract}
We show how to systematically apply the Faddeev-Jackiw symplectic method to General Relativity (GR) and to GR extensions. This provides a new coherent frame for Hamiltonian analyses of gravitational theories. The emphasis is on the classical dynamics, uncovering the constraints, the gauge transformations and the number of degrees of freedom; but the method results are also relevant for canonical quantization approaches. We illustrate the method with three applications: GR and to two Brans-Dicke cases (the standard case $\omega \not= - 3/2$ and the case with one less degree of freedom, $\omega = - 3/2$). We clarify subtleties of the symplectic approach and comment on previous symplectic-based Hamiltonian analyses of extended theories of gravity, pointing out that the present approach is systematic, complete and robust.
\end{abstract}


\maketitle

\section{Introduction}

General Relativity (GR) is currently the standard theory for gravitational phenomena. Considering phenomena at different scales, newer tests are being performed showing compatibility with GR \cite{Will:2014kxa, TheLIGOScientific:2017qsa, Rodrigues:2018duc, Collett:2018gpf}. Nonetheless, there are still different opens issues, anomalies and possible reasons for considering going beyond GR, and not only due to quantum gravity \cite{Capozziello:2010zz, Iorio:2014roa, 1107453984, Freedman:2017yms}.  Nowadays, the subtle aspects of GR, specially concerning its gauge symmetries, are much better understood than it was by the time it was proposed. Hence, sometimes it is useful to use the GR example to analyze new proposals. It must be remarked that such intuition developed from GR should be used with care, since  not all aspects of GR extensions are continuous extensions, one example being their numbers of degrees of freedom. General formalisms for dealing with the dynamical structure of a physical theory can disclose trustworthy and useful properties.

One of the fundamental theoretical developments of GR was its Hamiltonian formulation. The pioneering work of Arnowitt, Deser and Misner (see \cite{Arnowitt:1962hi, Deser:2015bia} for reviews), the ADM formalism, was important for various developments of GR, from canonical quantum gravity and the causal structure of GR to numerical GR. It is the most commonly used Hamiltonian formulation of GR, and it is based on a set of variables that has a clear dynamical meaning. It is surprising that, although the ADM formalism is known for decades, some of its fundamentals concerning gauge symmetries and other subtleties were discussed and elucidated only some years ago \cite{Kiriushcheva:2008fn, Frolov:2008sn, Mukherjee:2007yi, Kiriushcheva:2008sf, Pons:2009cz}. These issues were discussed within the standard formalism for constrained systems, the Dirac-Bergmann formalism (for reviews, see \cite{Dirac:1964:LQM, Gitman:1990qh, 0691037698}),

Here we consider another formalism for constrained systems, the symplectic formalism, more specifically the Faddeev-Jackiw formalism \cite{Faddeev:1988qp} with the Barcelos Neto-Wotzasek extension \cite{BarcelosNeto:1991kw,BarcelosNeto:1991ty}. The latter extension includes an iterative procedure to deal with the constraints. Our main goal in this work is to show how to apply this formalism to GR, elucidating some of its subtleties, and thus providing the means for applying the formalism to other GR-like theories. In the process, we clarify some general issues on this symplectic formalism approach, which concerns not only gravity theories (further details in the Conclusions). Apart from the application to GR, we also consider, to illustrate the formalism application to GR extensions, two Brans-Dicke theories, confirming some of the results presented in other papers, and showing explicitly, for the first time, the infinitesimal gauge transformations of these theories in terms of the ADM variables.

This paper is organized as follows: in the next section we review in detail the symplectic formalism and present some new comments, in particular a procedure to count the degrees of freedom entirely within the formalism. In section \ref{sec:GR} we apply apply the formalism to GR, together with the development of a suitable notation. Section \ref{sec:BD} applies the formalism to two Brans-Dicke cases, which serve as examples on how the formalism can be useful for extended gravity cases, display some minor new results on gauge symmetries, and uncover some general caveats on the symplectic formalism. Our conclusions are presented in Section \ref{sec:conclusions}. In the end, there are three appendices \ref{app:det}, \ref{app:timed} and \ref{app:eta2} presenting details of certain calculations.

\section{A review on symplectic methods for constrained systems} \label{sec:review}

We review in this section the symplectic method of Faddeev and Jackiw (FJ) \cite{Faddeev:1988qp, Jackiw:1993in}, and its extension as proposed by Barcelos Neto and Wotzasek (BW) \cite{BarcelosNeto:1991kw, BarcelosNeto:1991ty}. The latter combination is sometimes referred in the literature as modified FJ formalism, or simply as symplectic method.

\subsection{The  Faddeev-Jackiw method}

It is common to introduce the FJ method starting by presenting it in the context of a system of particles, but, since all the applications that will be performed here consider fields, this brief review will use the field notation from the start.

Let ${\cal L} = {\cal L}(\phi_a, \partial_\mu \phi_a, \partial_\mu \partial_\nu  \phi_a, \dots)$ be the Lagrangian density of a given theory that depends on the fields $\phi_a$ (with $a=1,2,...,A$) and on an arbitrary number of its derivatives,  where middle greek indices $\mu$ and  $\nu$  denote spacetime indices. For simplicity and clarity, we consider four dimensional spacetimes with metric signature $\begin{pmatrix} -1 & 1 & 1 & 1 \end{pmatrix}$. The first step of the FJ method is to write the Lagrangian density $\cal L$ as a function of certain  fields  $\xi^\alpha$, the symplectic fields, with $\alpha = 1,2,...,N$ (symplectic indices are denoted by initial greek indices), such that $\cal L$ depends  at most linearly  on the first time derivative of $\xi^\alpha$, $\dot \xi^\alpha$. For instance, for quadratic theories on the velocities $\dot \phi_a$, a common way to linearise the Lagrangian is to use the canonical momenta of the fields,
\begin{equation}
	\pi^a \equiv \frac{\partial {\cal L}}{\partial \dot \phi_a},
\end{equation}
yielding,
\begin{equation} \label{eq:generalSympVec}
	\xi^\alpha = \begin{pmatrix} \phi_a & \pi^a \end{pmatrix}.
\end{equation}

We remark that the index $a$ above may be understood as indexing  only the fields or the fields and their components. From now on we will only use $a$ to index the fields. Hence, $\phi_1$ and $\phi_2$ may respectively refer to a rank $p$ and a rank $q$ tensor, whose components would be indexed by internal indices of $\phi_1$ and $\phi_2$. Symplectic indices do not  have a straightforward relation to spacetime indices.

Independently on the original Lagrangian, and on the technique used to linearize it, to start the FJ method one should write the action in the following form,
\begin{equation}\label{action}
	S[\xi] = \int [a_\alpha(\xi) \dot \xi^\alpha  - {\cal V}(\xi)]d^4x,
\end{equation}
where $a_\alpha$ and $\cal V$ are respectively called the components of the canonical 1-form ($a = a_\alpha d\xi^\alpha$) and the potential. From this point onward, the dependence on the spatial derivatives will no longer be explicitly specified, hence $a_\alpha(\xi)$ in general cannot depend on $\dot \xi^\alpha$, but may depend on $\partial_i \xi^\alpha$ (with $i=1,2,3$) and higher order spatial derivatives.

Apart from surface terms, which will not be considered at this point, the field equations are found from the action variation,
\begin{equation} \label{eq:deltaS1}
	\delta S[\xi] = \int [\delta a_\alpha \dot \xi^\alpha +  a_\alpha \delta \dot \xi^\alpha  - \delta {\cal V}]d^4x.
\end{equation}

For a function $f=f(\xi)$, one can write the following useful relations \citep[see e.g.,][]{3642140890, 9783540591795},
\begin{equation}
	\delta f(x) = \int \frac{\delta f (x)}{\delta \xi^\alpha (x')} \delta \xi^\alpha (x')\, d^3x',
\end{equation}
\begin{equation}
	\dot f(x) = \int \frac{\delta f (x)}{\delta \xi^\alpha (x')} \dot\xi^\alpha (x')\, d^3x',
\end{equation}
\begin{equation}
	 \frac{d}{dt} \delta f =   \delta \frac{d}{dt}f = \delta \dot f,
\end{equation}
where the functional derivative satisfies,
\begin{eqnarray}
	 \frac{\delta \xi^\alpha(x)}{\delta \xi^\beta(x')} &=& \delta^\alpha_\beta \dd, \\[.1in]
	 \frac{\delta \partial_i \xi^\alpha(x)}{\delta \xi^\beta(x')} &=& \delta^\alpha_\beta \partial^x_i \dd = - \delta^\alpha_\beta \partial^{x'}_i \dd.
\end{eqnarray}
In the above, $x$ and $x'$ refer to different spacetime points, but with the same value for the $t$ coordinate (i.e., the derivatives are taken at equal time, $x'^0 = x^0 = t$).

Hence, apart from a surface term, eq.~(\ref{eq:deltaS1}) becomes,
\begin{eqnarray}
 \delta S &=& \int \left [ \frac{\delta a_\alpha(x)}{\delta \xi^\beta(x')} \dot \xi^\alpha(x) \delta \xi^\beta(x') - \frac{\delta a_\alpha(x)}{\delta \xi^\beta(x')} \dot \xi^\beta(x') \delta \xi^\alpha(x)  - \frac{\delta {\cal V}(x)}{\delta \xi^\beta (x')} \delta \xi^\beta(x')\right]d^3x\, d^3x'\, dt \nonumber \\[.1in]
  &=& \int \left [ f_{\alpha \beta}(x,x') \dot \xi^\beta(x')  - \frac{\delta {\cal V}(x')}{\delta \xi^\alpha (x)} \right]\delta \xi^\alpha(x) \, d^3x\, d^3x'\, dt, \label{eq:deltaS}
\end{eqnarray}
where the pre-symplectic matrix is defined as
\begin{equation}\label{presymplecticmatrix}
	f_{\alpha \beta} (x,x') \equiv \frac{\delta a_\beta(x')}{\delta \xi^\alpha(x)}  - \frac{\delta a_\alpha(x)}{\delta \xi^\beta(x')}.
\end{equation}

By demanding that $\delta S =0$ for an arbitrary variation $\delta \xi^\alpha$, the field equations can be written as
\begin{equation} \label{eq:symeq}
		\int f_{\alpha \beta}(x,x') \dot \xi^\beta (x') d^3x' = \frac{\delta V}{\delta \xi^\alpha(x)},
\end{equation}
with $V \equiv \int {\cal V}(x') d^3x'$. The procedure above is to be used independently on the existence of constraints (either already known and implemented by Lagrange multipliers, or yet to be discovered). If there are unknown constraints, they should be found from the field equations, and the constraints may be re-inserted into the action with the help of Lagrange multipliers, see Sec.~\ref{sec:review} for further details. Similarly to the Dirac formalism, one uses the assumption that the action contains all the relevant physical information, and even if the constraints are not explicit in the original action, they can be derived from it.

If $f_{\alpha \beta}$ has an inverse, then this matrix is called the symplectic matrix, and all the velocities $\dot \xi^\alpha$ can be derived from the field equations (\ref{eq:symeq}). Systems with this property are called non-singular. In this case, the dynamical evolutions of all the fields are uniquely determined, there are no gauge symmetries or constraints.

If $f_{\alpha \beta}$ is singular, then the pre-symplectic matrix has zero-modes (i.e., eigenvectors whose corresponding eigenvalues are zero). Let there be $M$ independent zero-modes denoted by $\nu_m^\alpha(x)$, then,
\begin{equation} \label{eq:zero}
	\int \nu^\alpha_m(x) f_{\alpha \beta}(x,x') d^3x =0,	
\end{equation}
where $m=1,2,...,M$. Therefore, from eq.~(\ref{eq:symeq}), one finds $M$ null relations given by
\begin{equation}\label{eq:consistCondition}
0 = \int  \nu^\alpha_m(x)\frac{\delta V}{\delta \xi^\alpha(x)}d^3x.
\end{equation}
The following special case can be commonly found in many examples of physical theories,
\begin{equation} \label{eq:zero2}
	\nu^\alpha_m(x) f_{\alpha \beta}(x,x') =0.
\end{equation}
If this particular case is true, then eq.~\eqref{eq:consistCondition} becomes
\begin{equation}\label{eq:consistCondition2}
0 =  \nu^\alpha_m(x)\frac{\delta V}{\delta \xi^\alpha(x)}.
\end{equation}

The eqs.~(\ref{eq:consistCondition}, \ref{eq:consistCondition2})  can either be trivial, if they simply lead to a known relation (i.e., $0=0$), or they can lead to new relations between the symplectic fields. The latter case implies the existence of constraints, which, for simplicity, we assume that they are all independent among themselves. If there are $M$ of such nontrivial relations, the system is said to have $M$ constraints given by
\begin{equation}\label{eq:constraint}
\Omega_{m}[\xi] \equiv \int \omega_{m}(\xi(x)) \, d^3x \equiv \int  \nu^\alpha_{m}(x)\frac{\delta V[\xi]}{\delta \xi^\alpha(x)}\, d^3x=0,
\end{equation}
with $m= 1,2,...,M$. If $\omega_m=0$, for all $m$, these equations are expected to determine a surface in the symplectic space, which is the constraint surface (see also Ref.~\cite{0691037698} for further details on the geometric interpretation).

The original FJ method \cite{Faddeev:1988qp, Jackiw:1993in} proposes to solve the constraints and use a Darboux transformation with the purpose of finding the true symplectic matrix. As was pointed out by Jackiw, ``{\it Of course there may be the technical obstacles to carrying out the above steps: solving the constraints may prove too difficult, constructing the Darboux transformation to canonical coordinates may not be possible}'' \cite{Jackiw:1993in}. One way to circumvent these issues is to simply abandon  this approach and move to the Dirac method. Another way is to continue within this approach and use the BW algorithm, which is briefly reviewed in the next subsection.

\subsection{The Barcelos Neto-Wotzasek (BW) extension of the FJ method} \label{sec:BW}

The BW algorithm \cite{BarcelosNeto:1991kw, BarcelosNeto:1991ty}  is an iterative procedure whose starting point is the Lagrangian density ${\cal L}^{(0)}$ linearized on the velocities, as implicitly given by the action (\ref{action}). This Lagrangian leads to the identification of the zeroth-order symplectic fields $\xi^{(0) \alpha}$, the components of the canonical 1-form $a_\alpha^{(0)}$ and the pre-symplect matrix $f^{(0)}_{\alpha \beta}$. This algorithm can be iteratively performed up to a certain step $r$ in which the symplectic matrix $f^{(r)}_{\alpha \beta}$ is found, without the need to eliminate the constraints or to find the appropriate Darboux transformation, as requested by the original FJ method.

The FJ method application to GR, without the BW algorithm and with the Darboux transformation, was performed in Refs.~\cite{Garriga:1997wz, Vitenti:2012cx}, with the purpose of finding the generalized (Dirac) brackets.

In general, to find the symplectic matrix (and hence the generalized brackets) using this method, it is necessary to fix the gauge. However,  this work aims to uncover the gauge generators and the constraints of a given gravitational theory, there should be no need to fix the gauge. We envisage to stop the BW iterative procedure at a certain step $r' \leq r$ in which no gauge fixing was done and all the constraints were found. This is also sufficient for a degree of freedom counting.

If the $f^{(0)}_{\alpha \beta}$ has zero-modes that lead to constraints $\omega_m^{(0)}$, as given by eq.~\eqref{eq:constraint}, the BW algorithm proposes to add this constraints to the kinetic part of ${\cal L}^{(0)}$, leading to \cite{BarcelosNeto:1991kw, BarcelosNeto:1991ty}
\begin{equation} \label{eq:L1}
	{\cal L}^{(1)}(\xi^{(0)}, \lambda^{(0)m}) \equiv {\cal L}^{(0)}(\xi^{(0)}) + \dot \lambda^{(0) m} \omega^{(0)}_m(\xi^{(0)}).
\end{equation}

In the above, ${\cal L}^{(1)}$ is dynamically equivalent to ${\cal L}^{(0)}$ since they only differ on the explicit imposition that the constraints should not evolve on time (i.e. $\dot \omega_m =0$). Also, ${\cal L}^{(1)}$ is already linear on the velocities, hence one can apply  the FJ method steps to the Lagrangian ${\cal L}^{(1)}$. To this end, one identifies $\left (\xi^{(1) \beta} \right ) = \left (\xi^{(0) \alpha}, \lambda^{(0)m}\right)$. If the index $\alpha$ associated to the zeroth iteration runs through 1 to $N$, and $m$ of the same iteration runs through 1 to $M$, then the $\beta$ of the first iteration runs from 1 to $N+M$. This procedure will lead to the first iteration pre-symplectic matrix $f^{(1)}_{\alpha \beta}$. If it still has zero-modes that yield new constraints $\omega^{(1)}_m$, the process is repeated by demanding that $\dot \omega^{(1)}_m = 0$, which leads to ${\cal L}^{(2)}$, defined analogously to ${\cal L}^{(1)}$ in eq.~\eqref{eq:L1}. The procedure stops once no new constraint is found.

In the BW algorithm, all the information on the constraints are implemented in the kinetic part; thus it is not hard to guess that there is no need to keep any constraints in the potential part after their implementation in the kinetic one \cite{BarcelosNeto:1991kw}. Indeed,  the discovered constraints can be iteratively eliminated from the potential (e.g., ${\cal V}^{(1)}\equiv {\cal V}^{(0)}|_{\omega_m^{(0)}=0}$). The sole purpose of this procedure is to ease the computations, while preserving the same dynamics on the constraint surface. This is a common and computationally useful procedure, but it is not mandatory.

\subsection{Gauge symmetries in the BW algorithm} \label{sec:gauge}

This subsection reviews the connection between gauge symmetries and zero-modes, as discussed in particular in Refs.~\cite{Montani:1992sy, Montani:1993hf, Montani:1998ip}. It also deals with issues related to field systems, and comments on a particular relevant case in which the rank of the pre-symplectic matrix becomes smaller on the constraint surface.  Various particular examples on uncovering gauge symmetries from the symplectic formalism within field systems can be found in the literature (e.g., \cite{Montani:1993hf, Wotzasek:1994ck, Neves:2003gz, Abreu:2013kpa}), but, to our knowledge, a general presentation about this case, highlighting its subtleties with respect to the particle system case, has not appeared before.

Within particle systems, the main result on the generators of gauge symmetries can be briefly stated as follows (see \cite{Montani:1992sy, Montani:1993hf, Montani:1998ip}): let a pre-symplectic structure, at some iteration of the BW algorithm, be degenerated in the constraint surface with $Z$ independent zero-modes which do not generate new constraints. Then, all these zero-modes will be associated to independent gauge transformations on the constraint surface. The relation between the gauge-related zero-modes and symplectic coordinate transformations is given by $\delta_\mtiny{G}\xi^{\alpha} = \nu^{\alpha}_{k} \varepsilon^{k}$, where $\delta_\mtiny{G}$ represents infinitesimal gauge transformations, the index $k$ is used to label the zero-modes, and $\{\varepsilon^{k}\}$ is a set of infinitesimal arbitrary parameters, one for each of the  zero-modes.

Indeed, if $\delta_{\mbox{\tiny G}} \xi^\alpha$ represents an infinitesimal gauge transformation on the symplectic fields and on the constraint surface, then, by definition,  $\delta_\mtiny{G} S = 0$ on the constraint surface [without using the field equations (\ref{eq:symeq})]. When it is relevant to stress that an equality holds on the constraint surface we use the ``weak equality'', introduced by Dirac, ``$\approx$''. Thus $\delta_\mtiny{G} S \approx 0$. From eq.~(\ref{eq:deltaS}), with $\delta \xi^\alpha = \delta_\mtiny{G} \xi^\alpha$, one gets
\begin{equation}
	\delta_\mtiny{G} S[\xi] = \int \left [ f_{\alpha \beta}(x,x') \dot \xi^\beta(x')  - \frac{\delta {\cal V}(x')}{\delta \xi^\alpha (x)} \right] \delta_\mtiny{G} \xi^\alpha(x) \, d^3x\, d^3x'\, dt  \, .
\end{equation}
Since, by hypothesis, $\delta_\mtiny{G} S \approx 0$ and as $\delta_\mtiny{G} \xi^\alpha$ is independent from the velocities $\dot \xi^\alpha$, one finds the two independent weak equalities,
\begin{eqnarray}
	\int  f_{\alpha \beta}(x,x')  \delta_\mtiny{G}\xi^\alpha(x)   d^3 x& \approx & 0 \, \label{eq:weakzeromode}\\[.1in]
	 \int \frac{\delta {\cal V}(x')}{\delta \xi^\alpha (x)}  \delta_\mtiny{G} \xi^\alpha(x) d^3 x & \approx & 0 \, . \label{eq:gauge1}
\end{eqnarray}
Therefore,  $\delta_\mtiny{G} \xi^\alpha$ is a gauge transformation on the constraint surface if and only if $\delta_\mtiny{G} \xi^\alpha$ is  a zero-mode of $f_{\alpha \beta}$ in the same surface and this zero-mode does not generate new constraints. The present work is not the first one to stress the importance of considering zero-modes on the constraint surface, see for instance Ref.~\cite{Wotzasek:1994ck}.

To conclude, we consider the issue of the general gauge generator. Let $\nu^\alpha_\varepsilon$ be the most general zero-mode of the pre-symplectic structure, and such that it satisfies (\ref{eq:gauge1}). The most general gauge transformation is therefore given by
\begin{equation} \label{eq:gaugeG}
	\delta_\mtiny{G} \xi^\alpha = \nu_\varepsilon^\alpha	\, .
\end{equation}
The relation between $\nu_\varepsilon^\alpha$ and $\nu^\alpha_k$ for particle  systems is given by $\nu_\varepsilon^\alpha = \nu_k^\alpha \varepsilon^k$, but this form is not in general valid for fields, due to the integration in eq.~\eqref{eq:zero}. In general, $\nu_\varepsilon$   depends on $Z$ arbitrary parameters $\varepsilon^k$ such that, for particular choices of $\varepsilon^k$, one can recover each of the particular linearly independent zero-modes $\nu^\alpha_k$.

\subsection{Number of degrees of freedom}

To our knowledge, the symplectic approach was not previously used to directly uncover the number of degrees of freedom (NDF). Here we present the general procedures to this end. For a review on the degrees of freedom counting, from the Dirac formalism, see Ref.~\cite{0691037698}. We remark that currently a number of gravity theories with nontrivial NDF is being considered. The best well known case is, probably, that of massive gravity and bigravity \cite{deRham:2014zqa}. The expected NDF for a massive spin-2 particle in four dimensional spacetime is five, but unless a very specific form for the mass term is chosen, one finds six degrees of freedom, the extra one being a ghost (e.g., \cite{Golovnev:2011aa, Huang:2013mha, Molaee:2017enn}). We also add that the NDF of massless spin-two and spin-zero fields are two and one, respectively; but a theory with these two fields needs not to have three degrees of freedom. Indeed, Brans-Dicke theory with $\omega = -3/2$ has an additional symmetry, a conformal invariance, which leaves the theory with 2 degrees of freedom. These results will be verified in the following sections within the formalism here proposed.

We start with the simplest case. Let ${\cal L}^{(0)}$ be a Lagrangian description of a theory with $N^{(0)}$ independent field components, this description is assumed to have no constraints or gauge symmetries. In this case, the zeroth step of the BW algorithm is already the final one, and the NDF in this case must be $\mbox{NDF} = N^{(0)}/2$.  Consider now the case where ${\cal L}^{(0)}$ describes a theory such that at the $k$-th iteration a total of $M$ independent constraints were found and $f^{(k)}$ is non-degenerate (i.e., $M$ is the total number of constraints and there are no gauge symmetries). Since each independent constraint can in principle be used to remove one of the independent field components, this theory has $\mbox{NDF} = (N^{(0)} - M)/2$.

In the previous example, in case $f^{(k)}$ has $G$ independent zero-modes that do not lead to new constraints, then $\mbox{NDF} = (N^{(0)} - M -  G)/2$. Indeed, one can always fix the gauge, and  for each independent zero-mode one should impose an independent condition on the original $N^{(0)}$ field components.

In the symplectic literature that uses the BW algorithm, it is common to find cases in which at some $k$-th iteration some of the field components are eliminated. Indeed, due to the process of eliminating the constraints from the potential (see Sec.~\ref{sec:BW}), eventually a field component that was present in the $(k-1)$-th iteration is no longer present in ${\cal L }^{(k)}$. If this happens for $E$ components, then $E$ independent field components will be eliminated along the algorithm, and one finds
\begin{equation} \label{eq:ndf}
	\mbox{NDF} = \frac 12 (N^{(0)} - M -  G -  E).
\end{equation}

For particle systems, the NDF should always be an integer number, and it must be compatible with Eq.~\eqref{eq:ndf}. Consider first that $G=E=0$. Indeed, the pre-symplectic matrix can only have an inverse at the $k$-th iteration if $N^{(k)}$ is even, since a square antisymmetric matrix with finite dimensions can only have an inverse if its dimension is even. For each independent constraint, one needs to insert a new Lagrange multiplier, hence for $M$ constraints (with no symplectic field elimination), one will have a symplectic vector with $N^{(0)} + M$ components. Assuming no gauge symmetry and no field elimination, if at this iteration the symplectic matrix is found, than necessarily $N^{(0)} + M$ is even. Consequently, $N^{(0)} + M - 2 M = N^{(0)} - M$ is also even, and the NDF is an integer.  This argument can be trivially extended to the case with gauge symmetry and field elimination, and one finds that the NDF computed from eq.~\eqref{eq:ndf} is always an integer for particle systems.

To conclude, since all iterations should generate Lagrangians that are dynamically equivalent among themselves, it must be possible to state  eq.~(\ref{eq:ndf}) as a function of $N^{(k)}$, being $k$ the iteration at which no new constraints are found. If at the $k$-th iteration $M$ constraints were found, then in ${\cal L}^{(k)}$ there should appear  $M$ field components that only appear once in ${\cal L}^{(k)}$ and with a time derivative. These are the Lagrange multipliers within the symplectic formalism. Since $N^{(k)}$ includes the number of Lagrange multipliers, which is always precisely $M$, we write
\begin{equation} \label{eq:ndfk}
	\mbox{NDF} = \frac 12 (N^{(k)} - 2 M -  G).
\end{equation}
Alternatively, eq.~(\ref{eq:ndfk}) can be found from eq.~(\ref{eq:ndf}) by using that   $N^{(k)} = N^{(0)} - E + M$.

\section{Application to General Relativity} \label{sec:GR}

\subsection{ADM variables and Lagrangian preparation}

Here we employ ADM variables (for reviews, see \cite{Hanson:1976cn, Wald:1984rg, Bojowald:2010qpa, Deser:2015bia}) and  we assume that spacetime is globally hyperbolic. Hence, it can be foliated by space-like hypersurfaces that can be parametrized by a scalar quantity $t$, these hypersurfaces are labeled $\Sigma_t$. The dynamics of GR from the ADM perspective can be seen as the evolution of the Riemmanian manifold ($\Sigma, h_{\mu \nu})$, where $h_{\mu \nu}$ is the induced tridimensional metric, and $\Sigma$ is a  three dimensional manifold whose metric changes along $t$. From the ADM perspective, the dynamical field is $h_{\mu \nu}$.

Using the ADM variables with an adapted coordinate system, the spacetime line element can be written as \cite{ Hanson:1976cn, Wald:1984rg, Bojowald:2010qpa}
\begin{equation} \label{eq:admline}
	ds^2  = - N^2 dt^2 + h_{ij}(dx^i + N^i dt)(dx^j + N^j dt)\, ,
\end{equation}
where the indices $i,j = 1,2,3$, the induced metric is $h_{ij}$, and $N$ and $N^i$ are respectively the lapse function and the shift vector.

Apart from surface terms, the action of GR reads,
\begin{equation} \label{eq:GR3+1}
	S[N, \bs{N}, \bs{h}] = \dfrac{1}{2 k }\int d^4x \, N\sqrt{h}[\3 R + K_{ij}K^{ij} - K^2],
\end{equation}
where $k  = 8 \pi G$, $\3 R$ and $K_{ij}$ are respectively the Ricci scalar and the extrinsic curvature of $\Sigma_t$, while $K = K_{i j} h^{i j} = K_i^i$. The fundamental fields of the theory are $N$, $N^i$ and $h_{ij}$. The extrinsic curvature, as a function of the fundamental fields, is
\begin{equation}
	K_{ij} = \frac{1}{2N} \left( \dot h_{ij} - D_i N_j - D_j N_i\right),
\end{equation}
where $D_i$ is the covariant derivative in $\Sigma$, and spatial indices are raised and lowered by $h_{ij}$.

In order to find the velocity-linearized Lagrangian, as in the action \eqref{action}, we employ the canonical momenta. Using that $S = \int {\cal L} \, d^4x$, the canonical  momenta are given by\footnote{If one uses the definition $\tilde \Pi_{ij} \equiv {\partial \cal L}/{\partial \dot h^{ij}}$, then the momenta will differ from eq.~\eqref{eq:PiRG} by a global sign, i.e. $\tilde \Pi_{ij} = - \Pi_{ij}$. We use $\Pi^{ij}$  since it is commonly adopted (e.g., \cite{ Hanson:1976cn, Wald:1984rg, Bojowald:2010qpa}).}
\begin{eqnarray}
	\Pi_{N} &\equiv& \frac{\partial \mathcal{L}}{\partial \dot{N}}=0 \, , \\[.2cm]
	\Pi_{i}&\equiv&\frac{\partial \mathcal{L}}{\partial \dot{N}^i}=0 \, ,  \\[.2cm]
	\Pi^{ij}&\equiv&\frac{\partial \mathcal{L}}{\partial \dot{h}_{ij}}= \frac{\sqrt{h}}{2 k } ( K^{ij} - K h^{ij}) \, . \label{eq:PiRG}
\end{eqnarray}

The last equation can be inverted,
\begin{equation}
	\dot{h}_{ij} = \frac{2 k  N}{\sqrt{h}} (2\Pi_{ij} - \Pi \, h_{ij}) + 2D_{(i}N_{j)},
\end{equation}
where $\Pi = \Pi^i_i$ and parenthesis indicate symmetrization, $A_{(i j)} = \frac 12 (A_{ij} + A_{ji})$.

It is now possible to write the velocity-linearized Lagrangian as
\begin{equation}
{\cal L}^{(0)} =
	\Pi^{ij} \dot{h}_{ij}  -  {\cal V}^{(0)} \,,
\end{equation}
where the potential reads
\begin{equation}
	{\cal V}^{(0)} =
		 \frac{2k  N}{\sqrt{h}} (\Pi_{ij} - \frac{1}{2} \Pi \,  h_{ij})\Pi^{ij} + 2\Pi_{ij}D^{i}N^{j} - \frac{1}{2k } N\sqrt{h} \3 R \, .
\end{equation}
Since the canonical momenta, with the usual conventions, were used to linearize the Lagrangian, the potential above is the canonical Hamiltonian of General Relativity (e.g.,  \cite{Wald:1984rg}).

\subsection{Notation conventions}

For the application to particular theories, it is convenient to introduce a clear and compact notation. Tensors in the tangent space of $\Sigma_t$ are denoted in boldface\footnote{To be more precise, we use boldface symbols as an ordered collection of components of tensors, not tensors themselves. That is, if $A$ is a second rank rank tensor on the tangent bundle of $\Sigma_t$, and if $\{ e^i\}$ is a basis of the tangent space, then $A = A_{ij} e^i \otimes e^j$.  The symbol $\bs{A}$ is a representation of $A$ in that basis, defined by  $\bs{A} \equiv (A_{ij})$. Also, one can use $\delta S/\delta \bs{A} \equiv (\delta S/  \delta A_{ij})$.}, while their individual components are specified with the letters $i,j,k, l,$ and $p$  [e.g., $\bs{N} = (N^i), \bs{h} = (h_{ij})$]. Spatial dependence on the coordinates  $x'^i$ are simply denoted by a prime in the corresponding field, $\Pi'^i = \Pi^i(\xp)$; for fields that depend on both  $\x$ and $\xp$  no prime is used. Tensors in the symplectic space are denoted by a boldface with a bar, and their components are written with the letters $\alpha, \beta, \gamma$ [e.g., $\xixi = (\xi^\alpha), \ff = (f_{\alpha \beta})$].  Each value of a symplectic index is associated to a field, not a to component of a field, hence one writes  $\xi^2 = \bs{N}$, or $\xi^{2_i} = N^i$. A sum in a symplectic index always imply that the corresponding internal ($\Sigma_t$) indices are summed as well, hence
\begin{equation}
	(\nunu \cdot \ff )_{\beta_{k l}}= \nu^\alpha f_{\alpha \beta_{k l}} = \sum_{\alpha, i, j} \nu^{\alpha_{ i j}} f_{\alpha_{i j} \beta_{k l}} \,.
\end{equation}

A $3 \times 3$ identity matrix is denoted by $\bs{1}$. We use $(\bs{1})^i_j = \delta^i_j$ and $(\bs{1})_j = \bs{1}_j = (\delta^i)_j$, i.e., $\bs{1}_j$ is the j-th line of the identity, in particular $\bs{1}_1 = ( 1 \;\; 0 \;\; 0 )$. There is  a type of identity element that is recurrent in the following computations, this lead us to introduce the  quantity $\bs{I}$ as follows,
\begin{equation}
	 \frac{\delta h_{i j}}{\delta h'_{kl}} = \delta_{(i}^k \delta_{j)}^l \dd \equiv I_{ij}^{kl},
\end{equation}
where the parenthesis on $i$ and $j$  indicate symmetrization.

The dot-product indicates that the maximum possible number of indices is being contracted, hence a dot-product between a rank 2 with a rank 4 tensor implies that two indices are being contracted. Some tensor contractions with omitted indices may seem at first not evident or ambiguous, but the dependence on the coordinates $\x$ or $\xp$ always sets which indices are being contracted.  Let $\bs{A} = (A^{i}), \bs{B} = (B^{i})$ and $\bs{C}=(C_{i})$, then
\begin{eqnarray}
	\bs{A} \cdot \frac{\delta \bs{C}'}{\delta \bs{B}} &= 	A^{i} \dfrac{\delta \bs{C}'}{\delta B^{i}}  & =  \left ( A^{i} \dfrac{\delta {C'_j}}{\delta B^{i}} \right ) \, , \nonumber \\[.1in]
	\bs{A} \cdot \frac{\delta \bs{C}}{\delta \bs{B}'} &= 	A^{i} \dfrac{\delta C_i}{\delta \bs{B}'} & =   \left( A^{i} \dfrac{\delta C_i}{\delta {B'^j}} \right )\, .
\end{eqnarray}

The iteration index $^{(i)}$ introduced by the Barcelos Neto-Wotzsek formalism is only displayed when necessary, commonly at the start of a new iteration.

\subsection{The zeroth iteration: finding all the constraints}
From the zeroth-iteration Lagrangian ${\cal L}^{(0)}$, one identifies the symplectic vector and the corresponding symplectic one-form,

\begin{eqnarray}\label{sympVect0GR}
   \xixi^{(0)  }  &=&
	\begin{pmatrix}
		N & \bs{N} & \bs{h} &  \bs{\Pi}
	\end{pmatrix} 	= \begin{pmatrix}
		N & N^i & h_{ij} &  \Pi^{ij}
	\end{pmatrix} \, , \\[.2cm]
		 {\aaa}^{(0)} &=&
	\begin{pmatrix}
		0 \; & \;\; 0 & \; \bs{\Pi} &  \; 0
	\end{pmatrix} = \begin{pmatrix}
		0 \; & \; \; 0_k & \;\Pi^{kl} &    \; 0_{k l}
	\end{pmatrix}\,.
\end{eqnarray}
From the above,
\begin{equation} 
 \frac{\delta a'_\beta}{\delta \xi^{\alpha}} = \delta_{\alpha}^4\delta_{\beta}^3 \frac{\delta \bs{\Pi'}}{\delta \bs{\Pi}}= 	\delta_{\alpha}^4\delta_{\beta}^3 \, \bs{I}\,.
\end{equation}
Thus, the pre-symplectic structure reads
\begin{equation} \label{eq:f0RGind}
 f_{\alpha \beta} =  \delta_{\alpha}^{4}\delta_{\beta}^{3} \, \bs{I}  - \delta_{\beta}^{4}\delta_{\alpha}^{3} \, \bs{I} \, .
\end{equation}

The matrix associated to the above pre-symplectic structure is clearly degenerate, and it is simple to find its zero-modes \eqref{eq:zero2}. Indeed:
\begin{equation}
 0=	(\nunu \cdot \ff)_\beta = \delta_{\beta}^{3} \, \bsnu^{4}\cdot \bs{I}  - \delta_{\beta}^{4}\, \bsnu^{3} \cdot \bs{I}  = \left ( \delta_{\beta}^{3} \, \bsnu^{4}  - \delta_{\beta}^{4}\, \bsnu^{3}  \right ) \dd \, .
\end{equation}

Any zero-mode must therefore satisfy\footnote{Since $\bsnu^3 \cdot \bs{I} = (\nu^{3_{ij}} I_{ij}^{kl})$ and  $I_{ij}^{kl} = I_{ji}^{kl} $, only the symmetric part of $\nu^{3_{i j}}$ and $\nu^{4_{i j}}$ must be null. Nonetheless, the anti-symmetric part has no information either on dynamics or on a relevant gauge symmetry, since all the rank 2 tensors on $\Sigma_t$ are symmetric; thus we take it to be zero.} $\bsnu^{3} = \bsnu^{4} = 0$. The components 1 and 2 of $\nu^\alpha$ were left arbitrary, thus the linearly independent zero-modes can be written as $\nu_\sigma^\alpha$, with $\sigma = 1, 2$, with $\nu^\alpha_1 = 0$, except for $\alpha = 1$; and  $\nu^\alpha_2 = 0$, except for $\alpha = 2$.

From the consistency condition (\ref{eq:consistCondition2}), for the zero-mode $\nunu_1$,
\begin{eqnarray}
	0 &=& \nu_{1}^{ \alpha} \dfrac{\delta V }{\delta \xi^{\alpha}}   =  \nu_{1}^{ 1} \dfrac{\delta V }{\delta N} \nonumber \\[.2cm]
	 &=& \nu_1^1 \left[\dfrac{2k }{\sqrt{h}} \left( \Pi_{ij} \Pi^{ij} - \dfrac{1}{2}  \Pi^{2} \right)
	 -  \dfrac{\sqrt{h}}{2 k }  \3 R \right]\;.
\end{eqnarray}
The expression above leads to the Hamiltonian constraint,
\begin{eqnarray}
	\omega_0 = \dfrac{2 k }{\sqrt{h}} \left( \Pi_{ij} \Pi^{ij} - \dfrac{1}{2} \Pi^{2} \right) -  \dfrac{\sqrt{h}}{2 k } \3 R\;.
\end{eqnarray}

The consistence condition for  the  second zero-mode $\nunu_2$ reads
\begin{eqnarray}
	0 = \nu_{2}^{\alpha} \dfrac{\delta V }{\delta \xi^{\alpha}}  =  \nu_{2}^{2_i} \dfrac{\delta V }{\delta N^{i}}   = -  2 \nu_{2}^{2_i} D^{j}\Pi_{j i}  \;.
\end{eqnarray}
Thus, we find the three diffeomorphism constraints,
\begin{eqnarray}
	\omega_{i} = - 2  D^{j} \Pi_{j i}\;.
\end{eqnarray}
The name for these constraints  are somewhat misleading within the symplectic formalism, since gauge symmetries are not generated by special types of constraints, but by special types of zero-modes. In the next iteration the gauge symmetries will be found.

With the above, the zeroth iteration is complete. Since constraints were found, one proceeds to the next iteration, thus
\begin{eqnarray}
	{\mathcal{L}}^{(1)} = \dot{h}_{ij} \Pi^{ij} + \dot{\lambda}^0 \omega_{0} + \dot{\lambda}^i \omega_{i}\, .
\end{eqnarray}
where the known constraints were eliminated from the potential, leading to
\begin{equation} \label{eq:V1RG}
	{\cal V}^{(1)} =0\, .
\end{equation}
Since the potential is null, it is possible to guarantee at this point that there are no new constraints to be uncovered. The infinitesimal gauge transformations are found in the next subsection.\footnote{From the perspective of the Dirac formalism, if the Hamiltonian can be written as a linear combination of the known constraints no new constraint will be found. In this case $\dot \omega_m \approx \{\omega_m, H\} \approx \sum_{m'} \lambda_{m'} \{\omega_m, \omega_{m'}\}$, therefore either all the constraints are of first class, and thus all Lagrange multipliers will not be determined; or some of the constraints will be of second class, leading to the determination of some Lagrange multipliers \cite{0691037698}. In both cases, no new constraints appear.  }

\subsection{The first iteration: uncovering the gauge symmetries} \label{sec:firsti}

The symplectic vector and the symplectic one-form of the first iteration are identified as
\begin{eqnarray}
	{\xixi}^{ (1)}  &=&
	\begin{pmatrix}
		\bs{h} & \bs{\Pi} & \lambda^0 & \; \bs{\lambda}
	\end{pmatrix}	=
	\begin{pmatrix}
		h_{ij} & \Pi^{ij} & \lambda^0 & \lambda^i
	\end{pmatrix} \, ,   \nonumber \\[.1cm]
	{ \aaa}^{(1)} &=&
	\begin{pmatrix}
		\bs{\Pi} &  0  & \; \omega_0 \; & \bsomega
	\end{pmatrix} =
	\begin{pmatrix}
		\Pi^{kl} & 0_{kl}    &  \omega_0 \; & \omega_{k}
	\end{pmatrix}\,.
\end{eqnarray}
All the fields that do not appear in ${\cal L}^{(1)}$  were omitted from $\xixi^{(1)}$. With the above,
\begin{equation}\label{eq:axifirst}
	\frac{\delta a'_\beta}{\delta \xi^{\alpha}} = \delta_\alpha^1\delta_\beta^3 \frac{\delta \omega'_0}{\delta \bs{h}}+ \delta_\alpha^1\delta_\beta^4 \frac{\delta \bsomega'}{\delta \bs{h}} +\delta_\alpha^2\delta_\beta^1 \, \bs{I}+  \delta_\alpha^2\delta_\beta^3 \frac{\delta \omega'_0}{\delta \bs{\Pi}} +  \delta_\alpha^2\delta_\beta^4 \frac{\delta \bs{\omega'}}{\delta \bs{\Pi}} \, .
\end{equation}

The pre-symplectic structure can be found by properly anti-symmetrizing eq.~\eqref{eq:axifirst}. It can be represented in matrix form by associating each $\alpha$ value to a line and each $\beta$ value to a column, thus
\begin{eqnarray}\label{preSympMatrixRG}
	 \ff =
	\begin{pmatrix}
		0 & - \bs{I}  & \dfrac{\delta \omega'_0}{\delta \bs{h}} & \dfrac{\delta \bsomega'}{\delta \bs{h}} \\[0.2cm]
		\bs{I}  & 0 & \dfrac{\delta \omega'_0}{\delta \bs{\Pi}} & \dfrac{\delta \bsomega'}{\delta \bs{\Pi}}  \\[0.2cm]
		-\dfrac{\delta \omega_0}{\delta \bs{h}'} & -\dfrac{\delta \omega_0}{\delta \bs{\Pi}'} & 0 & 0 \\[0.2cm]
		-\dfrac{\delta \bsomega}{\delta \bs{h}'}  & -\dfrac{\delta \bsomega}{\delta \bs{\Pi}'} & 0 & 0
	\end{pmatrix}\;.
\end{eqnarray}
One can explicitly compute the determinant of the above matrix and show that, in the complete symplectic space, it is not degenerated. Nonetheless, in the constraint surface, it is degenerated, as shown in  Appendix \ref{app:det}. Therefore, we need to find zero-modes such that  $\int \nunu \cdot \ff \, d^3 x \approx 0$, see also eq.~(\ref{eq:weakzeromode}). We shall first consider  the following two components,
\begin{eqnarray}
	0 \approx \int (\nunu \cdot \ff)_3 d^3 x&=& \int \left( \bsnu^{1} \cdot \dfrac{\delta \omega'_0}{\delta \bs{h}} + \bsnu^2 \cdot \dfrac{\delta \omega'_0}{\delta \bs{\Pi}} \right ) d^3x \, , \label{ff3} \\
	0 \approx \int (\nunu \cdot \ff)_{4} d^3x&=& \int \left( \bsnu^{1} \cdot \dfrac{\delta \bsomega'}{\delta \bs{h}} + \bsnu^2 \cdot \dfrac{\delta \bsomega'}{\delta \bs{\Pi}} \right)d^3x\,  \label{ff4}.
\end{eqnarray}
It is not hard to realize that for particular choices of $\bsnu^{1}$ and $\bsnu^{2}$ the above structure can be expressed as Poisson brackets among fields. Since the zero-mode can only depend on the spatial coordinate $x$, it cannot depend on variational derivatives of the functions\footnote{Since $\delta \omega_0(x)/ \omega_i(x')$ depends on both $x$ and $x'$.} $\omega_0$ or $\omega_i$, but it can depend on derivatives of the functional $\Omega_\mu = \int \omega_\mu d^3x$ (with $\mu=0,1,2,3$). Then, making use of the well known Dirac algebra (see e.g., \cite{Dirac:1951zz, Castellani:1981us, Mukherjee:2007yi}),
\begin{eqnarray}
		\{\omega_0, \omega'_0 \} &=& (\omega^i + \omega'^i ) \partial_i \dd,	\nonumber \\
		\{\omega_i, \omega'_0 \} &=& \omega_0 \partial_i\dd, \label{eq:constraintalgebra} \\
		\{\omega_i, \omega'_j \} &=& \omega'_i \partial_j\dd + \omega_j \partial_i\dd \, . \nonumber
\end{eqnarray}
one can get the four $\mu$ indexed solutions to eqs.~(\ref{ff3}, \ref{ff4}):
\begin{eqnarray}
	\bsnu^1_\mu = \frac{\delta \Omega_\mu}{\delta \bs{\Pi}} \, , \;\;\; \bsnu^2_\mu = -\frac{\delta \Omega_\mu}{\delta \bs{h}}\, .
\end{eqnarray}
From the knowledge of the components 1 and 2, it is easy to derive the remaining components. The four zero-modes read
 \begin{eqnarray}
	\nunu_{0} &=&
	\begin{pmatrix}
		 \frac{\delta \Omega_0}{\delta \bs{\Pi} } & - \frac{\delta \Omega_0}{\delta \bs{h}} & -1 & \; 0
	\end{pmatrix}, \label{modoZeroRG11} \\[.1in]
	\nunu_{p} &=&
	\begin{pmatrix}
		 \frac{\delta \Omega_p}{\delta \bs{\Pi} } & - \frac{\delta \Omega_p}{\delta \bs{h}} & \; 0 \; & - \bs{1}_p
	\end{pmatrix}. \label{modoZeroRG12}
\end{eqnarray}

One can directly verify that $\nunu_\mu$ are indeed zero-modes on the constraint surface (which we call ``weak zero-modes''),
\begin{equation}
	 \int (\nunu_\mu \cdot {\ff})_\beta \, d^3x  = \delta^3_\beta \{\omega'_0, \Omega_\mu\} + \delta^4_\beta \{\bsomega', \Omega_\mu\}	\approx 0\,.
\end{equation}

In order to uncover the gauge symmetries, the weak zero-modes must be generalized by introducing an arbitrary infinitesimal field. From the algebra \eqref{eq:constraintalgebra}, it is easy to see that, for any $\varepsilon^\nu$,
\begin{equation}
\int 	\{ \omega_\mu , \varepsilon'^\nu \omega'_\nu\} d^3 x' = \int \left( \varepsilon'^\nu \{ \omega_\mu,   \omega'_\nu \} +  \{ \omega_\mu ,   \varepsilon'^\nu  \} \omega'_\nu \right) d^3 x' \approx 0.
\end{equation}
This implies that the zero-modes (\ref{modoZeroRG11}, \ref{modoZeroRG12}) can be generalized to depend on an arbitrary infinitesimal vector field $\varepsilon^\mu$, which is achieved by replacing $\Omega_\mu$ with $ \Omega_\varepsilon \equiv \int \varepsilon^\mu(x) \omega_\mu(x) d^3x$. Indeed, one can write the following general weak zero-mode,
\begin{equation}\label{modoZeroRG05ep}
	{\nunu}_{{\varepsilon}} =
	\begin{pmatrix}
		 \frac{\delta \Omega_\varepsilon}{\delta \bs{\Pi}} & - \frac{\delta \Omega_\varepsilon}{\delta \bs{h}} & -\varepsilon^0 & \boldsymbol{\varepsilon}
	\end{pmatrix}.
\end{equation}
Each of the four zero-modes (\ref{modoZeroRG11}, \ref{modoZeroRG12}) are found from $\nunu_{{\varepsilon}}$ by a particular choice of $\varepsilon^\mu$.

The infinitesimal gauge transformations of $h_{ij}$ in the constraint surface are found from the first component of $\nunu_{{\varepsilon}}$,
\begin{eqnarray}
	\delta_\mtiny{G}h_{ij}  & \approx &   \frac{\delta \Omega_{\varepsilon}}{\delta \Pi^{ij}} \nonumber \\
		&\approx &  \int \left( \varepsilon'^0 \frac{\delta \omega'_0}{\delta \Pi^{ij}} + \varepsilon'^k \frac{\delta \omega'_k}{\delta \Pi^{ij}} \right ) d^3x' \nonumber \\
		&\approx &  \int \left [ \varepsilon'^0 \frac{4 k}{\sqrt{ h'}}\left( \Pi'_{ij} - \frac 12 \Pi' h'_{ij}\right)\dd  - 2 \varepsilon'_k\frac{\delta D'_l \Pi'^{lk}}{\delta \Pi^{ij}}  \right]\,  d^3x' \nonumber \\[.1in]
		&\approx &  2 k \, \varepsilon^0 \, K_{i j} + 2 D_{(i} \varepsilon_{j)} \, .
\end{eqnarray}

The above result, within the Dirac formalism, can also be found in\footnote{There is a sign difference in the term proportional to $\varepsilon^0$, but this is due to their definition of the extrinsic curvature, which differs from ours by a global sign.} Ref.~\cite{Mukherjee:2007yi} (more precisely,  their eq.~(44)). The term $2 D_{(i} \varepsilon_{j)}$ in the above equation is a Lie derivative, hence it is immediate to interpret it as coordinate transformations on $\Sigma_t$ alone. The $\varepsilon^0$ term refers to changes along the normal direction of $\Sigma_t$. The above expression is the correct one. Equivalent results can be obtained in a similar way for the phase space variable $\Pi^{ij}$, although the gauge changes normal to $\Sigma_t$ are much more cumbersome in this case. For further details on the interpretation of gauge transformations of GR within the ADM variables, see Refs.~\cite{Castellani:1981us, Mukherjee:2007yi, Kiriushcheva:2008fn, Shestakova:2011ek, Frolov:2008sn, Kiriushcheva:2008sf}.

\subsection{Degrees of freedom counting} \label{sec:dof}

In order to count the number of degrees of freedom in General Relativity, we proceed as follows, using eq.~(\ref{eq:ndf}): the total number of field components in $\xixi^{(0)}$ is 16 (one scalar, one vector and two symmetric tensors); from those original components, four were eliminated ($N$, $N^i$); four constraint components were found ($\omega_0, \omega_i$); and, in the last iteration, four independent  zero-modes  were found $\nunu_\mu$. Hence, as expected, there are $\frac 12 (16-4-4-4) = 2$ degrees of freedom in GR.


\section{Applications to two Brans-Dicke  Theory cases} \label{sec:BD}

\subsection{The Brans-Dicke action and momenta}

The Brans-Dicke action with a potential reads \cite{Brans:1961sx, Fujii:2003pa, Capozziello:2010zz}
\begin{equation}
\label{action27}
	S[g,\phi] = \frac 1{2 k } \int \left (  \phi R   - \frac{\omega}{\phi} \partial_\mu \phi \partial^\mu \phi - P(\phi) \right ) \sqrt{-g} \, d^4x \,,
\end{equation}
with $\omega$ being a constant and  $P(\phi)$  the scalar field potential. This action is  not the most general scalar-tensor gravity, but it is sufficiently simple and interesting. It includes the original Brans-Dicke proposal ($P(\phi)=0$) \cite{Brans:1961sx}, it is dual to the metric $f(R)$ gravity if $\omega =0$, and it is dual to the Palatini $f(R)$ gravity if $\omega = -3/2$ \cite{Capozziello:2010zz}. This action is known for having  three degrees of freedom if  $\omega \not= -3/2$, and 2 degrees of freedom if $\omega = -3/2$ \cite{Olmo:2011fh}. In vacuum, the theory with $\omega = - 3/2$ has a conformal symmetry and it is possible to map its solutions to the GR ones \cite{Ferraris:1992dx}. The corresponding action is a reformulation of GR in vacuum, and it is a good example on how to apply the formalism, as it has some relevant subtleties that are further developed in the end of this section and in Appendix \ref{app:timed}.

Similarly to the case of general relativity, we use ADM variables,  as in eq. \eqref{eq:admline}, see also \cite{Olmo:2011fh,Moon:2013ska},
\begin{eqnarray}
		S[g,\phi] &=& \frac{1}{2 k } \int \left\{ N \phi[\3 R  + K_{i j} K^{ij} - K^2]+ 2 D_i N D^i \phi - 2 K(\dot \phi - N^j D_j \phi)  +\right. \nonumber \\
		& & \left. + \frac{\omega}{N\phi} [ (\dot \phi - N^i D_i\phi)^2 - N^2 D_i \phi D^i \phi] - N P(\phi)\right \} \sqrt h \, d^4x \, .
\end{eqnarray}

The momenta read\footnote{Regarding eq.~\eqref{eq:piphi}, there is a  misprint on sign of $K$ in the corresponding equation in Ref.~\cite{Olmo:2011fh}. } (see also Ref.~\cite{Olmo:2011fh})
\begin{eqnarray}
	\Pi_N &=& \frac{\partial {\cal L}}{\partial \dot N} = 0 \, , \\[0.2cm]	
	\Pi_i &=& \frac{\partial {\cal L}}{\partial \dot N^i} = 0 \, , \\	[0.2cm]	
	\Pi^{ij} &=& \frac{\partial {\cal L}}{\partial \dot h_{ij}} =  \frac{\sqrt{h}}{2 k } \left [\phi (K^{ij} - K h^{ij}) - \frac{1}{N} h^{i j} (\dot \phi - N^k D_k \phi)) \right]\, , \label{eq:piBD}\\[0.2cm]		
	\Pi_\phi &=& \frac{\partial {\cal L}}{\partial \dot \phi} = \frac{\sqrt h}{ k } \left ( -  K + \frac{ \omega}{N \phi} (\dot \phi - N^iD_i \phi)\right ) \, . \label{eq:piphi}
\end{eqnarray}

In order to linearize the Lagrangian with respect to the velocities, it is useful to note that
\begin{equation}\label{eq:PiPiphiphi}
	\Pi - \phi \Pi_\phi = \frac{\sqrt h}{2 k }\frac{(3 + 2 \omega)}{N} \left ( N^i D_i \phi - \dot \phi \right ) \, .
\end{equation}

\subsection{Brans-Dicke with $\omega \not= -3/2$}

The Lagrangian density linearized in the velocities reads \cite{Zhang:2011vg},
\begin{eqnarray} \label{eq:Lbrans}
	{\cal L}^{(0)} &=& \dot{h}_{ij} \Pi^{ij} + \dot{\phi} \Pi_{\phi}
	+  \dfrac{\sqrt{h}}{2 k }  N \left ( \phi \3 R - \dfrac{2 k ^2}{ h\phi} \left( 2 \Pi^{ij} \Pi_{ij} - \Pi^{2} \right) - \dfrac{\omega}{\phi} D_{i}\phi D^{i} \phi - \nonumber \right. \\[.1in]
	&& \left. - 2 D_iD^i \phi  - \dfrac{2 k ^2}{h \phi (3+2\omega)} \left( \Pi - \phi \Pi_{\phi} \right)^{2} - 2 P(\phi) \right)  	+ N^{j} \left(  2 D^{i}\Pi_{ij} - \Pi_{\phi} D_{j} \phi \right) \;.
\end{eqnarray}

For the symplectic vector, we select the most economical form (neglecting any field that does not appear in ${\cal L}$), hence
\begin{eqnarray}\label{sympVectBD}
	{ \xixi}^{(0)  }  &=&
	\begin{pmatrix}
		N & N^i & h_{ij} &  \Pi^{ij} & \phi & \Pi_\phi
	\end{pmatrix} \, , \\[.2cm]
	{\aaa}^{(0)} &=&
	\begin{pmatrix}
		0 \; & \; 0_k & \;\Pi^{kl} &   0_{k l} & \Pi_\phi & 0 \;\;
	\end{pmatrix}\,.
\end{eqnarray}
From these vectors we compute
\begin{equation}
 f_{\alpha \beta} 	=  \delta_{\alpha}^4\delta_{\beta}^3 \, I^{k l}_{i j} + \delta^6_\alpha \delta^5_\beta \, \dd - \delta_{\alpha}^3 \delta_{\beta}^4 \, I_{k l}^{i j} - \delta^5_\alpha \delta^6_\beta \, \dd \, .
\end{equation}

The zero-modes of $\ff$ are vectors $\nunu$ such that all their components are null, except for $\nunu^1$ and $\nunu^2$, which are arbitrary. One finds that there are four independent zero-modes, which lead to the following four constraints,
\begin{eqnarray}
	\omega_0 &=&  \dfrac{\sqrt{h}}{2 k }  \left ( -\phi \3 R + \dfrac{2 k ^2}{ h\phi} \left( 2 \Pi^{ij} \Pi_{ij} - \Pi^{2} \right) + \dfrac{\omega}{\phi} D_{i}\phi D^{i} \phi  + 2 D_iD^i \phi  + \right. \nonumber \\[.1in]
	&& \left. + \dfrac{2 k ^2}{h \phi (3+2\omega)} \left( \Pi - \phi \Pi_{\phi} \right)^{2} + 2 P(\phi) \right)\, , \\[.2in]
\omega_{j} &=&   - 2 D^{i}\Pi_{ij} + \Pi_{\phi} D_{j} \phi \, .
\end{eqnarray}

The first iteration Lagrangian, with the constraints removed from the symplectic potential, reads
\begin{eqnarray}
	{\mathcal{L}}^{(1)} = \dot{h}_{ij} \Pi^{ij} + \dot \phi \Pi_\phi + \dot{\lambda}^0 \omega_{0} + \dot{\lambda}^i \omega_{i}\, .
\end{eqnarray}
 The first-iteration potential is zero, therefore  $\omega_0$ and $\omega_i$ are all the constraints of the theory.

The symplectic vector and the symplectic one-form of the first iteration are taken to be
\begin{eqnarray}
	{ \xixi}^{ (1)}  &=&
	\begin{pmatrix}
		h_{ij} & \Pi^{ij} &  \; \phi & \Pi_\phi & \lambda^0 & \lambda^i
	\end{pmatrix} \nonumber \\[.1cm]
	{\aaa}^{(1)} &=&
	\begin{pmatrix}
		\Pi^{kl} & \; 0_{kl}     &  \Pi_\phi & 0 &  \omega_0 \; & \omega_{k}
	\end{pmatrix}\,.
\end{eqnarray}

The pre-symplectic structure reads,
\begin{eqnarray}\label{preSympMatrixBD}
	 {\ff}  =
	\begin{pmatrix}
		0 & - \bs{I}  &  0 & 0 & \dfrac{\delta \omega'_0}{\delta \bs{h}} & \dfrac{\delta \bsomega'}{\delta \bs{h}} \\[0.2cm]
		\bs{I}  & 0  & 0 & 0 & \dfrac{\delta \omega'_0}{\delta \bs{\Pi}} & \dfrac{\delta \bsomega'}{\delta \bs{\Pi}}  \\[0.2cm]
		0 & 0 & 0 & -\ddc & \dfrac{\delta \omega'_0}{\delta \phi} & \dfrac{\delta \bsomega'}{\delta \phi} \\[.2cm]
		0 & 0& \ddc & 0 & \dfrac{\delta \omega'_0}{\delta \Pi_\phi} & \dfrac{\delta \bsomega'}{\delta \Pi_\phi}   \\[.2in]
		-\dfrac{\delta \omega_0}{\delta \bs{h}'} & -\dfrac{\delta \omega_0}{\delta \bs{\Pi}'} & - \dfrac{\delta \omega_0}{\delta \phi'}  & -\dfrac{\delta \omega_0}{\delta \Pi'_\phi} & 0 & 0\\[0.2cm]
		-\dfrac{\delta \bsomega}{\delta \bs{h}'}  & -\dfrac{\delta \bsomega}{\delta \bs{\Pi}'} & -\dfrac{\delta \bsomega}{\delta \phi'}& - \dfrac{\delta \bsomega}{\delta \Pi'_\phi}& 0 & 0
	\end{pmatrix}\;,
\end{eqnarray}
with $\ddc \equiv \dd$. The above $\ff$ can be promptly  seen as an extension of eq.~(\ref{preSympMatrixRG}), and it only has zero-modes on the constraint surface. Following analogous steps, the weak zero-modes must be given by
\begin{eqnarray}
	\nunu_0 &= &
	\begin{pmatrix}
		 \dfrac{\delta \Omega_0}{\delta \bs{\Pi}} & - \dfrac{\delta \Omega_0}{\delta \bs{h}} & \dfrac{\delta \Omega_0}{\delta \Pi_\phi} & - \dfrac{\delta \Omega_0}{\delta \phi} & \; -1 \; & \; 0	
	\end{pmatrix} \, ,	 \\[.2in]
	\nunu_p &= &
	\begin{pmatrix}
		 \dfrac{\delta \Omega_p}{\delta \bs{\Pi}} & - \dfrac{\delta \Omega_p}{\delta \bs{h}} &\dfrac{\delta \Omega_p}{\delta \Pi_\phi} & - \dfrac{\delta \Omega_p}{\delta \phi} & \;\; 0  & \;- \bs{1}_p
	\end{pmatrix} \,.	
\end{eqnarray}

In order to verify that the above vectors satisfy $\int \nunu \cdot \ff \,  d^3x \approx 0$, one needs to use the Brans-Dicke constraint algebra. It is not necessary to use explicitly the algebra details, the only relevant relation is that, for Brans-Dicke, likewise in GR, $\{\Omega_\mu, \omega_\nu \} \approx 0$ (see for instance Ref.~\cite{Zhang:2011vg}). Conversely, if one was not aware of the Brans-Dicke algebra, but knew that its action is a scalar, the latter condition would need to be true (otherwise, there would be no gauge symmetry on the constraint surface, and hence no diffeomorphism invariance).

The Brans-Dicke theory with $\omega \not= -3/2$ has three degrees of freedom, indeed, from eq.~(\ref{eq:ndfk}): (18 - 2 $\times$ 4 - 4)/2 = 3 degrees of freedom.

The gauge transformations are  found from the most general zero-mode. It is an extension of eq.~\eqref{modoZeroRG05ep}, and it reads
\begin{equation} \label{nuepsilonBD1}
{\nunu}_{\bs \varepsilon} =
	\begin{pmatrix}
			 \dfrac{\delta \Omega_\varepsilon}{\delta \bs{\Pi}} & - \dfrac{\delta \Omega_\varepsilon}{\delta \bs{h}} &  \dfrac{\delta \Omega_\varepsilon}{\delta \Pi_\phi} & - \dfrac{\delta \Omega_\varepsilon}{\delta \phi} & -\varepsilon^0 & -\boldsymbol{\varepsilon}
	\end{pmatrix}.
\end{equation}

The gauge transformation for the dynamical fields $\bs{h}$ and $\phi$ are respectively found from the first and the third component of $\nunu_\varepsilon$, namely,
\begin{eqnarray}
\delta_\mtiny{G} h_{ij} & = & \frac{\delta \Omega_\varepsilon}{\delta \Pi^{ij}}  \nonumber \\
	&= &  2k  \varepsilon^0K_{ij} + \frac{2k \epsilon^0}{\sqrt{h}\phi(3+2\omega)} (\Pi - \phi\Pi_{\phi})h_{ij} + 2 D_{(i} \varepsilon_{j)} \nonumber \\
	& = & 2k  \varepsilon^0 K_{ij} + \frac{\varepsilon^0}{\phi N}  (N^l D_l \phi - \dot \phi)h_{ij} + 2 D_{(i} \varepsilon_{j)}\, , \label{gaugeBD1h}\\[.1in]
\delta_\mtiny{G} \phi &= &  \frac{\delta \Omega_\varepsilon}{\delta \Pi_\phi}  \nonumber \\
	&=&  \frac{\varepsilon^0}{\phi N}  (N^l D_l \phi - \dot \phi) \phi  + \varepsilon^i D_i \phi \, . \label{gaugeBD1phi}
\end{eqnarray}

Gauge transformations for $\Pi_{ij}$ and $\Pi_{\phi}$ can be obtained in a similar way. To our knowledge, the above gauge transformations have not yet explicitly appeared in the literature. For the Brans-Dicke gauge transformations with a different set of variables, displaying a $SU(2)$ gauge symmetry, see Ref.~\cite{Zhang:2011vg}. Similarly to the GR case, the diffeomorphism invariance on the $\Sigma_t$ surface can be promptly spotted as the last terms of eqs.~(\ref{gaugeBD1h}, \ref{gaugeBD1phi}), which are both Lie derivatives.

\subsection{Brans-Dicke with $\omega = - 3/2$} \label{sec:BD2}

The case with $w=-3/2$ should have one degree of freedom less than the previous $ w \not= 3/2$ Brans-Dicke theory, see Refs.~\cite{Capozziello:2010zz, Ferraris:1992dx}). Indeed, from eq.~\eqref{eq:PiPiphiphi} one can promptly see that a new constraint emerges,
\begin{eqnarray}
\eta_1 \equiv \Pi - \Pi_\phi \phi  = 0.
\end{eqnarray}
In Dirac language, this is a primary constraint, as it comes directly from the momenta definition, hence either one solves the constraint eliminating one of the fields, or the constraint must be inserted in the Lagrangian. We add it to the Lagrangian following the same formalism rules we have been using before, that is, adding the time derivative of the constraint (or the time derivative of the Lagrange multiplier). We remark that adding this constraint without any time derivative is incompatible with the symplectic formalism we are adopting, as it will be commented in the Conclusions and demonstrated in the Appendix \ref{app:timed}.

The linearized Lagrangian on the velocities reads (see also  \cite{Olmo:2011fh, Zhang:2011vg}),
\begin{eqnarray} \label{eq:Lbrans32}
	{\cal L}^{(0)} &=& \dot{h}_{ij} \Pi^{ij}  + \dot \phi \Pi_\phi +  \dot \zeta_1 \eta_1 + N^{j} \left(  2 D^{i}\Pi_{ij} - \Pi_{\phi} D_{j} \phi \right) + \nonumber \\
	&& +N \dfrac{\sqrt{h} }{2 k }   \left ( \phi \3 R - \dfrac{2 k ^2}{ h\phi} \left( 2 \Pi^{ij} \Pi_{ij} - \Pi^{2} \right) + \dfrac{3}{2 \phi} D_{i}\phi D^{i} \phi  - 2 D_iD^i \phi   - 2 P(\phi) \right) \, .
\end{eqnarray}

The symplectic vector and the canonical 1-form are written as
\begin{eqnarray}\label{sympVect0GR32}
{ \xixi}^{(0)  }  &=&
\begin{pmatrix}
N & N^i & h_{ij} &  \Pi^{ij} & \phi & \Pi_\phi & \zeta_1
\end{pmatrix} \, , \\[.2cm]
{ \aaa}^{(0)} &=&
\begin{pmatrix}
0 & \; 0_k & \Pi^{kl} &   0_{k l} & \Pi_\phi & 0 \; & \; \eta_1 \;
\end{pmatrix}\,.
\end{eqnarray}
From the above we get,
\begin{equation}
\frac{\delta { a}'_\beta}{\delta {\xi}^{\alpha}}  = 	\delta_{\alpha}^4\delta_{\beta}^3 \, \bs{I} + \delta^6_\alpha \delta^5_\beta \dd +  \left(\delta_\alpha^3 \bs{\Pi} + \delta_\alpha^4 \bs{h} - \delta_\alpha^5 \Pi_\phi - \delta_\alpha^6 \phi \right )\delta_\beta^7 \dd\, .
\end{equation}
Hence, to satisfy  $(\nunu \cdot \ff)_{\beta} = 0 $, one finds the following equations (each one for a different value of $\beta$),
\begin{eqnarray}
0 &=& \bsnu^4 - \nu^7 \,  \bs{\Pi}  \, ,\\
0 &=& \bsnu^3 +  \nu^7 \,  \bs{h} \, ,  \\
0 &=&\nu^6 + \nu^7 \, \Pi_\phi    ,\\
0 &=&- \nu^5+  \nu^7 \, \phi  \, ,\\
0 &=&  \bsnu^{3} \cdot \bs{\Pi} +  \bsnu^4 \cdot \bs{h}  - \nu^5 \, \Pi_\phi  - \nu^6  \, \phi \,  .
\end{eqnarray}
The first four equations fix four components of $\nunu$ as functions of $\nu^7$. The fifth equation is not independent, it can be found from the previous four. Hence, there are  three linearly independent zero-modes, which are denoted by $\nu^\alpha_\sigma$, with $\sigma = 1,2,3$ and read
\begin{eqnarray}
	\nunu_1 &=& \begin{pmatrix} 1 & 0 & \;\;  0 & \;  0 & \; 0 & \;\; \, 0 & \;\;\; 0\end{pmatrix}, \\
	\nunu_{2_p} &=& \begin{pmatrix} 0 & \bs{1}_p & \; 0 & \;  0 & \;  0 & \;\; 0 & \;\;\; 0\end{pmatrix}, \\
	\nunu_3 &=& \begin{pmatrix} 0 & 0 & - \bs{h} & \bs{\Pi} & \phi & -\Pi_\phi & 1\end{pmatrix} . \label{eq:nu3BD}
\end{eqnarray}

The zero-modes ${\nunu}_1$ and ${\nunu}_2$, from the consistency equation, yield respectively the following constraints:
\begin{eqnarray}
\omega_0 &=& \dfrac{\sqrt{h} }{2 k }   \left ( \phi \3 R - \dfrac{2 k ^2}{ h\phi} \left( 2 \Pi^{ij} \Pi_{ij} - \Pi^{2} \right) + \dfrac{3}{2 \phi} D_{i}\phi D^{i} \phi  - 2 D_iD^i \phi   - 2 P(\phi) \right) \, , \label{omega0RG} \\[.1in]
\omega_i &=&    - 2 D^{k}\Pi_{ki} + \Pi_{\phi} D_{i} \phi\, . \label{omegaiRG}
\end{eqnarray}

The zero-mode $\bsnu_3$ leads to the constraint
\begin{eqnarray}
0 &=&   \int \left( -h_{ij} \frac{\delta}{\delta h_{ij}} + \Pi^{ij} \frac{\delta}{\delta \Pi^{ij}} + \phi \frac{\delta}{\delta \phi}- \Pi_\phi \frac{\delta}{\delta \Pi_\phi} \right)  V d^3x \nonumber \\
&\approx & \int  N\sqrt{h}( -2P + \phi \partial_\phi P ) d^3x\, . \label{findingeta2constraint}
\end{eqnarray}
Further details on the computation above can be found in Appendix \ref{app:eta2}. From the above, we identify a new constraint\footnote{This constraint can also be found directly from the field equations (e.g., \cite{Capozziello:2010zz}).}
\begin{equation} \label{eta2}
\eta_2 =  -2P + \phi \partial_\phi P \, .
\end{equation}
This constraint has a simple solution,
\begin{equation} \label{Psol}
	P = \lambda \phi^2,
\end{equation}
where $\lambda$ is a dimensionless constant. Note that $\lambda$ is not a mass scale. Indeed, by making the field redefinition $\phi = \varphi^2$ in order to put the kinetic term of the scalar field in canonical form in Lagrangian \eqref{action27}, one gets the usually well known conformally invariant potential $\lambda \varphi ^4$, as it is the only one with a dimensionless coupling constant. No other scalar field potential can be made compatible with conformal invariance, as it will necessarily introduce a dimensional coupling constant leading to a fundamental scale in the theory. This is an important remark that will be used later on.

There is no reason to keep the constraint $\eta_2$ in explicit form further, since there is no other possible development besides the solution above. Thus, we eliminate this constraint. The scalar potential $P$ is, from now on, taken to be given by eq.~\eqref{Psol}.

The first iteration Lagrangian, with the constraints removed from the symplectic potential, reads
\begin{eqnarray}
	{\mathcal{L}}^{(1)} = \dot{h}_{ij} \Pi^{ij} + \dot \phi \Pi_\phi + \dot{\lambda}^0 \omega_{0} + \dot{\lambda}^i \omega_{i} + \dot \zeta_1 \eta_1 \, .
\end{eqnarray}
Since the potential has disappeared at this iteration, no new constraints can be found. The symplectic vector and the canonical 1-form are written as
\begin{eqnarray}
	{ \xixi}^{(1)  }  &=&
	\begin{pmatrix}
		h_{ij} &  \Pi^{ij} & \phi & \Pi_\phi & \zeta_1 & \lambda^0 & \lambda^i
	\end{pmatrix} \, , \\[.2cm]
	{ \aaa}^{(1)} &=&
	\begin{pmatrix}
		\Pi^{kl} &   0_{k l} & \Pi_\phi & 0 \; & \eta_1  & \omega_{0} & \omega_{k}   \;
	\end{pmatrix}\,.
\end{eqnarray}
The pre-symplectic structure reads,
\begin{eqnarray}\label{preSympMatrixBDmod}
		{ \ff}  =
	\begin{pmatrix}
		0 & - \bs{I}  &  0 & 0&  \bs{\Pi}  & \dfrac{\delta \omega'_0}{\delta \bs{ h}} & \dfrac{\delta \bsomega'}{\delta \bs{h}}   \\[0.2cm]
		\bs{I}  & 0  & 0 & 0 &   \bs{h} & \dfrac{\delta \omega'_0}{\delta \bs{\Pi}} & \dfrac{\delta \bsomega'}{\delta \bs{\Pi}}   \\[0.2cm]
		0 & 0 & 0 & -\ddc &   -\Pi_{\phi}  & \dfrac{\delta \omega'_0}{\delta \phi} & \dfrac{\delta \bsomega'}{\delta \phi}   \\[.2cm]
		0 & 0& \ddc & 0 &  -\phi & \dfrac{\delta \omega'_0}{\delta \Pi_\phi} & \dfrac{\delta \bsomega'}{\delta \Pi_\phi}   \\[.2in]
		- \bs{\Pi}' & - \bs{h}' & \Pi_{\phi}' & \phi' & 0 & 0 & 0  \\[0.2cm]
		-\dfrac{\delta \omega_0}{\delta \bs{h}'} & -\dfrac{\delta \omega_0}{\delta \bs{\Pi}'} & - \dfrac{\delta \omega_0}{\delta \phi'}  & -\dfrac{\delta \omega_0}{\delta \Pi'_\phi} & 0 & 0 & 0  \\[0.2cm]
		-\dfrac{\delta \bsomega}{\delta \bs{h}'}  & -\dfrac{\delta \bsomega}{\delta \bs{\Pi}'} & -\dfrac{\delta \bsomega}{\delta \phi'}& - \dfrac{\delta \bsomega}{\delta \Pi'_\phi}& 0 & 0 & 0  \\[0.2cm]
	\end{pmatrix}\;. \nonumber
\end{eqnarray}

Since this model comes from an action that is invariant under coordinate transformations, it is already known that it must posses a zero-mode, parametrized by $\varepsilon^\mu$, that extends eq.~\eqref{modoZeroRG05ep} (see also eq.~\eqref{nuepsilonBD1}). Indeed, using that $\{ \Omega_\mu, \omega'_\nu\} \approx 0 $, it is straightforward to verify that the following vector is a zero-mode on the constraint surface,
\begin{equation} \label{nuepsilonBD2}
{\nunu}_{\bs \varepsilon} =
	\begin{pmatrix}
			 \frac{\delta \Omega_\varepsilon}{\delta \bs{\Pi}} & - \frac{\delta \Omega_\varepsilon}{\delta \bs{h}} &  \frac{\delta \Omega_\varepsilon}{\delta \Pi_\phi} & - \frac{\delta \Omega_\varepsilon}{\delta \phi} & 0 & -\varepsilon^0 & -\boldsymbol{\varepsilon}
	\end{pmatrix}.
\end{equation}

Since the constraint surface is  spanned by five independent constraints, there may be up to five linearly independent zero-modes. Considering the first five columns of $\ff$, one can find another zero-mode candidate [see also eq.~(\ref{eq:nu3BD})],
\begin{equation}
	{\nunu}_{\eta} = \eta
	\begin{pmatrix}
		- \bs{h}  & \bs{\Pi}  &  \phi & - \Pi_{\phi} & 1 & 0 & 0
	\end{pmatrix}\;,
\end{equation}
where $\eta$ in an infinitesimal arbitrary field. To verify that ${\nunu}_{\eta}$ is indeed a weak zero-mode of $\ff$, one uses the same computations already derived in Appendice \ref{app:eta2}.

To count the number of degrees of freedom for the Brans-Dicke theory with $\omega = -3/2$, we use eq.~(\ref{eq:ndfk}):  (19 - 2 $\times$ 5 - 5)/2 = 2 degrees of freedom.

The gauge transformation for  $\bs{h}$ and $\phi$ now depend on five parameters ($\varepsilon^\mu$ and $\eta$) and are respectively found from the first and the third components of the most general zero-mode, $\nunu_\varepsilon + \nunu_\eta$, therefore,
\begin{eqnarray}
\delta_\mtiny{G} h_{ij} & = & \frac{\delta \Omega_\varepsilon}{\delta \Pi^{ij}} - \eta h_{ij}  \nonumber \\
	& = &  2k  \varepsilon^0K_{ij} +  2 D_{(i} \varepsilon_{j)} - \eta h_{ij} \, . \\[.2in]
\delta_\mtiny{G} \phi & = &  \frac{\delta \Omega_\varepsilon}{\delta \Pi_\phi}  + \eta \phi \nonumber \\
	&=&   \varepsilon^i D_i \phi  + \eta \phi \, . 
\end{eqnarray}
This shows  that in the present Brans-Dicke case, besides the gauge symmetry related to coordinate transformations, parametrized by $\varepsilon^\mu$, there appears a conformal gauge transformation parametrized by $\eta$.

\section{Conclusions} \label{sec:conclusions}

Here we have shown how to apply the iterative symplectic formalism \cite{Faddeev:1988qp, BarcelosNeto:1991kw, BarcelosNeto:1991ty, Montani:1992sy}  to GR and to two cases of the Brans-Dicke theory. In the process, we have clarified issues in the general formalism and opened the way to  applications to other extended formulations of gravity. Below, we stress and comment on some results of this work:

\noindent
{\it (1) Degrees of freedom counting.} In Sec.~\ref{sec:dof} we introduced a method, completely within the symplectic formalism, to count the degrees of freedom.

\noindent
{\it (2) Generalized zero-mode and gauge transformations.} We introduced the most general zero-mode parametrized by an arbitrary field. For particle systems, a simple multiplication of the zero-mode by an arbitrary parameter $\varepsilon(t)$ is sufficient, but for the fields it is shown that this process depends in general on an integration of the arbitrary parameter \eqref{modoZeroRG05ep}, which leads to the introduction of $\Omega_\varepsilon$.

\noindent
{\it (3) Weak zero-modes and diffeomorphism invariance.}  In the symplectic formalism,  eigenvectors with eigenvalues that are a linear combination of the constraints need to be considered among the zero-modes of the symplectic matrix \cite{Wotzasek:1994ck}. These kind of zero-modes we named weak zero-modes, in reference to the weak equality introduced by Dirac. To our knowledge, this is the first work to point its relation to diffeomorphism invariance, and to explicitly derive the zero-modes that generate gauge symmetries in GR, which can be parametrized by $\varepsilon^\mu$ [see eq.~\eqref{modoZeroRG05ep}].

\noindent
{\it (4) On the symplectic approach of Escalante and collaborators.} In Refs.~\cite{Escalante:2015aea,Escalante:2016qky,Escalante:2017fzh} another approach to the iterative symplectic algorithm can be found. In their approach, some columns of the pre-symplectic structure were ignored in the process of finding the zero-modes. The reason why it is hard to find the zero-modes of the complete matrix \eqref{preSympMatrixRG} was  clarified in this work and just stated above: there are no such zero-modes, only weak zero-modes. In some cases, their approach can lead to the correct results, but there is no  proof that one can simply ignore some columns and always find the correct answer. From the Dirac-Bergmann formalism perspective, this is analogous of stating that some constraints are first-class constraints without verifying the Poisson brackets among these  constraints. It may work, but for each system there should be a good explanation on why it is not necessary to verify all the Poisson brackets. Anyway, here we provided a detailed proof that the symplectic formalism can be applied to GR and to extended theories of gravity, and there is no need to neglect part of the pre-symplectic matrix. Also, as shown in Sec.~\ref{sec:firsti}, the last columns of the pre-symplectic matrix can be useful for uncovering the (weak) zero-modes. Hence, even in the cases where it is possible to ignore the last columns, to do so is not necessarily the fastest procedure.

\noindent
{\it (5) On the importance of the order of the constraints and the time derivatives of the Lagrange multipliers.} Reference~\cite{Toms:2015lza}, while commenting on the symplectic formalism, states that time derivatives on the Lagrangian multipliers can be used, but are innocuous, since Lagrange multipliers are arbitrary. In a broad sense there is  truth in this remark, but we stress that within the symplectic formalism (and probably any Hamiltonian formalism), this statement should be understood with great care. As explicitly shown in Appendix \ref{app:timed}, changing a Lagrangian multiplier to its time derivative does change the physics emerging from the symplectic formalism: it can change the amount of constraints that are found, leading to physically non-equivalent results. This issue appears in particular in  the application presented in Sec.~\ref{sec:BD2}.

\noindent
{\it (6) Notation and arbitrary rank tensors.}  The notation introduced here, which associates each symplectic index value to a field, not to a field component, can be promptly employed to theories with arbitrary rank tensors.

We expect that this method, and extensions based on it, will prove fruitful for the analysis of specific systems within GR or for extended theories of gravity. The formalism here presented also provides a parallel framework that can work as a cross-check for the results derived from other approaches.

\acknowledgements
DCR and MG thank \'Alefe Freire de Almeida for relevant discussions during the beginning of this work. DCR thanks Raju Roychowdhury for comments on a draft version of this work and Cl\'ovis Wotzasek for long ago discussions on the symplectic formalism that proved useful to this work. DCR thanks FAPES (Brazil) and CNPq (Brazil) for partial  support. NPN would like to thank CNPq of Brazil for financial support PQ-IB number 309073/2017-0.

\appendix

\section{The determinant of the pre-symplectic structure of GR} \label{app:det}
To compute the determinant of \eqref{preSympMatrixRG}, we use the same technique of \cite{Ellicott:1990up, Toms:2015lza}, which starts from the observation that, for a given square matrix $M$ that can be subdivided in four blocks, one writes
\begin{equation}
	\det M = \det \begin{pmatrix}
		A & B \\
		C & D
	\end{pmatrix}	= \det (D - C A^{-1} B) \, \det A \, .
\end{equation}
In the above, it is assumed that $A$ and $D$ are square matrices and that $\det A \not= 0$. The matrices $B$ and $C$ need not to be square matrices.

We set
\begin{eqnarray}
 && A = \begin{pmatrix}
 	0 & - \bs{I}\\
 	\bs{I} & 0
 \end{pmatrix} , 
 \;\;\; B = \begin{pmatrix}
 	\dfrac{\delta \omega'_0}{\delta \bs{h}} & \dfrac{\delta \bsomega'}{\delta \bs{h}} \\[.1in]
 	\dfrac{\delta \omega'_0}{\delta \bs{\Pi}} & \dfrac{\delta \bsomega'}{\delta \bs{\Pi}}
 \end{pmatrix}	\, , \nonumber\\[.1in]
 && C  = \begin{pmatrix}
 	-\dfrac{\delta \omega_0}{\delta \bs{h}'} & -\dfrac{\delta \omega_0}{\delta \bs{\Pi}'} \\[.1in]
 	- \dfrac{\delta \bsomega}{\delta \bs{h}'} & - \dfrac{\delta \bsomega}{\delta \bs{\Pi}'}
 \end{pmatrix}	\, , 
  D  = \begin{pmatrix}
 	0 & 0 \\
 	0 & 0
 \end{pmatrix}	\, ,
\end{eqnarray}
that is, the matrix $M$ is the pre-symplectic matrix $\ff$ from eq.~(\ref{preSympMatrixRG}).  Using the above definitions for the blocks $A, B, C, D$ and the algebra \eqref{eq:constraintalgebra}, we find,
\begin{eqnarray}
(C A^{-1} B)_{a b} (x, x''')&=&	\int C_{a c}(x,x') (A^{-1})^{c d}(x',x'') B_{d b}(x'', x''') \, d^3x' d^3x'' \nonumber \\
	&=& \int \left [\begin{pmatrix}
			\dfrac{\delta \bs{\omega}_0}{\delta \bs{\Pi}''} & -\dfrac{\delta \omega_0}{\delta \bs{h}''} \\[.2cm]
			\dfrac{\delta \bsomega}{\delta \bs{\Pi}''} & - \dfrac{\delta \bsomega}{\delta \bs{h}''}
		\end{pmatrix}
		\begin{pmatrix}
 			\dfrac{\delta \omega'''_0}{\delta \bs{h}''} & \dfrac{\delta \bsomega'''}{\delta \bs{h}''} \\[.2cm]
			\dfrac{\delta \omega'''_0}{\delta \bs{\Pi}''} & \dfrac{\delta \bsomega'''}{\delta \bs{\Pi}''}
		\end{pmatrix} \right ]_{ab}
		d^3 x'' \nonumber \\[.4cm]
	&=&  \begin{pmatrix}
			\{\omega_0''', \omega_0 \} & \{ \bsomega''', \omega_0 \} \\
			\{\omega_0''', \bsomega \} & \{\bsomega''', \bsomega \}
		\end{pmatrix}_{ab}  \label{CAB}\\[.2in]
	&\approx& 0 \nonumber \, .
\end{eqnarray}
In the full symplectic space $\det \ff \not=0$, while on the constraint surface, $\det \ff = 0$. This implies that $\ff$ has no zero-modes in the full symplectic space, but it has at least one zero-mode in the constraint surface.

\section{On the proper use of the time derivative of constraints} \label{app:timed}

In the Lagrangian of  eq.~\eqref{eq:Lbrans32}, the constraint $\eta_1$ appears in the term $\eta_1 \dot \zeta_1 $, where $\zeta_1$ is the Lagrangian multiplier. This constraint $\eta_1$ was found directly from the definition of momenta, that is, in the Dirac nomenclature it would be a primary constraint. Contrary to the other primary constraints that appear in this paper, this is the single one that does not imply an obvious field elimination. For instance, the primary constraint $\Pi_{N} = 0$ found in Sec.~\ref{sec:GR} simply leads to the elimination of $\Pi_N$. One could as well solve $\eta_1$, say eliminating $\Pi_\phi$ in favor of the other quantities, but in this case there are more than one possible field elimination, and breaking such symmetry may lead to technical difficulties and inconveniences. Hence, this is a primary constraint that is useful to be kept. The symplectic principles tell us that constraints, once found, should be added to the Lagrangian with the time derivative of the Lagrange multiplier. This is the procedure followed in Sec.~\ref{sec:BD2}. On the other hand, one may think that there is no harm in inserting a term without time derivatives,  as for instance suggested in a comment of Ref.~\cite{Toms:2015lza}. We explore this path here.

By using $\eta_1 \zeta_1$ in Lagrangian \eqref{eq:Lbrans32}, $\xixi$ and $\aaa$ become
\begin{eqnarray}
{ \xixi}^{(0)  }  &=&
\begin{pmatrix}
N & N^i & h_{ij} &  \Pi^{ij} & \phi & \Pi_\phi & \zeta_1
\end{pmatrix} \, , \\[.2cm]
{ \aaa}^{(0)} &=&
\begin{pmatrix}
0 & \; 0_k & \Pi^{kl} &   0_{k l} & \Pi_\phi & 0 \; & \; \; 0 \;
\end{pmatrix}\,,
\end{eqnarray}
that is, $\eta_1$ does not appear in $\aaa$. Consequently,
\begin{equation}
\frac{\delta { a}'_\beta}{\delta {\xi}^{\alpha}}  = 	\delta_{\alpha}^4\delta_{\beta}^3 \, \bs{I} + \delta^6_\alpha \delta^5_\beta \dd \, .
\end{equation}
The pre-symplectic matrix has now the following zero-modes,
\begin{eqnarray}
	\nunu_1 &=& \begin{pmatrix} 1 & 0 & \;  0 & \;  0 & \; 0 & \; 0 & \; 0\end{pmatrix}, \\
	\nunu_{2_p} &=& \begin{pmatrix} 0 & \bs{1}_p &  0 & \;  0 & \; 0 & \; 0 & \; 0\end{pmatrix}, \\
	\nunu_3 &=& \begin{pmatrix} 0 & \; 0 & \;  0 & \; 0 & \; 0 & \; 0  & \; 1\end{pmatrix} .
\end{eqnarray}
The two first zero-modes lead to the same constraints (\ref{omega0RG}, \ref{omegaiRG}), while $\eta_1$ is found as a constraint from the last zero-mode. This may seem to show the equivalence between the approaches, since although $\eta_1$ was put in the potential part at first, it was in the end found as a constraint. However, in the procedure of inserting the time derivative of the Lagrangian multiplier from the start, a new constraint, $\eta_2$, is found at the zeroth-iteration. The main problem is that $\eta_2$ is neither found at the zeroth-iteration, nor at any other iteration: at the first iteration the full potential becomes null (it is just a linear combination of constraints), and therefore it is impossible to find any new constraints.

In conclusion, if $\eta_1 \zeta_1$ is inserted in the Lagrangian instead of $\eta_1 \dot \zeta_1$, one is not following the symplectic formalism  (since all constraints should be inserted in the Lagrangian with time derivatives) and these different procedures are not physically equivalent. In this case, the constraint $\eta_2$ is not obtained, and it is a physically fundamental constraint saying that the unique non-trivial potential compatible with conformal invariance is the $\lambda\varphi ^4$ potential.

Similarly to the previous case, since $N$ and $N^i$ can be promptly seen to be Lagrange multipliers in GR and Brans-Dicke theories, one can consider the possibility of using a shortcut such that  $N$ and $N^i$ are replaced by  $\dot \lambda^0$ and $\dot \lambda^i$ before the first iterative step. This procedure can work in some cases, leading to correct and faster results, but there is no guarantee that it will always work. Indeed, for the Brans-Dicke case with $\omega = -3/2$ this procedure misses the constraint  $\eta_2$, which is a fundamental constraint for the self-consistency of the theory.

\section{The determination of the $\eta_2$ constraint for Brans-Dicke with $\omega = - 3/2$} \label{app:eta2}

The $\eta_2$ constraint is derived from the following relation [see eq.~\eqref{findingeta2constraint}],
\begin{equation} \label{operatoreta2}
	 \int \left( -h_{ij} \frac{\delta}{\delta h_{ij}} + \Pi^{ij} \frac{\delta}{\delta \Pi^{ij}} + \phi \frac{\delta}{\delta \phi}- \Pi_\phi \frac{\delta}{\delta \Pi_\phi} \right) V \, d^3x= 0 \, .
\end{equation}

The potential $V$ comes from the Lagrangian \eqref{eq:Lbrans32}, and it reads,
\begin{eqnarray}
	V &=& - \int \left \{ N \dfrac{\sqrt{h} }{2 k }   \left [ \phi \3 R - \dfrac{2 k ^2}{ h\phi} \left( 2 \Pi^{ij} \Pi_{ij} - \Pi^{2} \right) + \dfrac{3}{2 \phi} D_{i}\phi D^{i} \phi  - 2 D_iD^i \phi   - 2 P(\phi) \right] + \right. \nonumber \\[.2in]
	&&  + N^{j} \left(  2 D^{i}\Pi_{ij} - \Pi_{\phi} D_{j} \phi \right)  \bigg \} \,  d^3x  = \int ( - N\omega_0 + N^i \omega_i) d^3x \, .
\end{eqnarray}
The constraints $\omega_0$ and $\omega_i$ are defined in eqs.~(\ref{omega0RG}, \ref{omegaiRG}).

To ease the computation of eq.~\eqref{operatoreta2}, we subdivide $V$ into seven terms ($V = \sum_{i=1}^7 V_i$). These terms are explicitly stated below,
\begin{eqnarray}
	V_1 = -  \int \frac{\sqrt{h}}{2 \kappa} N \phi \3 R \, d^3x \, ,& \;\;\;\;   & V_2 = \kappa \int \frac{N}{\sqrt{h} \phi} \left( 2 \Pi^{i j} \Pi_{i j} - \Pi^2\right) d^3x \nonumber \\[.2in]
	V_3 = - \frac {1}{2 \kappa} \int \frac{3 N \sqrt{h} }{2 \phi} D^i \phi D_i \phi \, d^3 x \, , &  \; \; \; \;  &  V_4 = - \frac 1 \kappa \int \sqrt h D_i N D^i \phi \, d^3x\, , \nonumber \\[.2in]
	 V_5 = \int \frac{ N \sqrt h}{\kappa} P(\phi) \, d^3 x \, , & \;\;\;\; & V_6 = \int N^i \Pi_\phi D_i \phi d^3 x \, , \\[.2in]
	 V_7 = 2 \int \Pi^{k m}D_k N^l h_{l m} \, d^3x\, . & \;\;\;\; & \nonumber
\end{eqnarray}

The application of each of the variations that appear in eq.~\eqref{operatoreta2} to each of the terms of $V$ is displayed in Table \ref{tab}. Putting the individual results together, one finds,
\begin{eqnarray}
	0 &= &\frac 12 \sum_{i=1}^5 V_i - 2 V_5 + \int \frac{N \sqrt{h}}{\kappa} \phi \partial_\phi P(\phi) d^3x  \nonumber \\
&\approx &\int \frac{ N \sqrt h}{\kappa} [	-2 P(\phi) + \phi \partial_\phi P(\phi)  ]d^3x \, ,
\end{eqnarray}
where it was used that $\sum_{i=1}^5 V_i \propto \omega_0$. Therefore, one finds the constraint $\eta_2$, as given by eq.~\eqref{eta2}.

\setlength{\tabcolsep}{12pt} 
\begin{table}[ht]
\caption{Results on the applications of the operators in eq.~\eqref{operatoreta2} to the seven terms of $V$.}\label{tab}
\begin{tabular}{l|cccc}
   & $\displaystyle \int h_{ij} \dfrac{\delta}{\delta h_{ij}}$  & $ \displaystyle \int \Pi^{ij} \dfrac{\delta}{\delta \Pi^{ij}}$ &  $ \displaystyle  \int \phi \dfrac{\delta}{\delta \phi}$ & $ \displaystyle  \int \Pi_\phi \dfrac{\delta}{\delta \Pi_\phi}$ \\
  \hline
  $V_1$&  $ V_1 / 2$ & 0 & $V_1$ & 0  \\
  $V_2$&  $V_2/2$ & $2 V_2$ & $- V_2$ & 0 \\
  $V_3$&  $V_3/2$ & 0 & $V_3$ & 0\\
  $V_4$& $V_4/2$ & 0 & $V_4$ & 0 \\
  $V_5$& $3 V_5/2$ & 0 & $\int N \sqrt h \phi \partial_\phi P(\phi) d^3 x / \kappa$ & 0 \\
  $V_6$& 0 & 0 & $V_6$ & $V_6$ \\
  $V_7$& $V_7$& $V_7$ & 0 & 0
\end{tabular}
\end{table}

\bibliographystyle{apsrev4-1}
\bibliography{bibdavi2016c}

\end{document}